\documentclass[aps,prb,twocolumn,superscriptaddress]{revtex4-1}

\usepackage[utf8]{inputenc}

\usepackage[american]{babel}

\usepackage{amsfonts}
\usepackage{amsmath}
\usepackage{amssymb}
\usepackage{textcomp}
\usepackage{color,graphicx}
\usepackage{xcolor}
\usepackage{siunitx}
\usepackage{microtype}
\usepackage{hyperref}

\DeclareSIUnit\pJ{\pico\joule}
\DeclareSIUnit\fJ{\femto\joule}

\usepackage[autostyle,english=american]{csquotes}

\newcommand{\subfigid}[1]{#1}

\newcommand{\oureqref}[1]{\ref{#1}}

\begin{document}

\title{Effective electron temperature measurement using \\ time-resolved anti-Stokes photoluminescence}

\author{Thomas Jollans}
\affiliation{Huygens--Kamerlingh Onnes Laboratory, Leiden University, Leiden, The Netherlands}
\author{Martín Caldarola}
\affiliation{Kavli Institute of Nanoscience Delft, Department of Quantum Nanoscience, 
Delft University of Technology, The Netherlands}
\affiliation{Kavli Institute of Nanoscience Delft, Department of Bionanoscience, 
Delft University of Technology, The Netherlands}
\author{Yonatan Sivan}
\affiliation{School of Electrical and Computer Engineering, Ben-Gurion University of the Negev, Beer-Sheva, Israel}
\author{Michel Orrit}
\email{Corresponding author: orrit@physics.leidenuniv.nl}
\affiliation{Huygens--Kamerlingh Onnes Laboratory, Leiden University, Leiden, The Netherlands}

\date{\today}

\begin{abstract}
Anti-Stokes photoluminescence of metal nanoparticles, in which emitted photons have a higher energy than the incident photons, is an indicator of the temperature prevalent within a nanoparticle. Previous work has shown how to extract the temperature from a gold nanoparticle under continuous-wave monochromatic illumination. We extend the technique to pulsed illumination and introduce pump-probe anti-Stokes spectroscopy. This new technique enables us not only to measure an effective electron temperature in a gold nanoparticle (\SI{\sim e3}{\kelvin} under our conditions), but also to measure ultrafast dynamics of a pulse-excited electron population, through its effect on the photoluminescence, with sub-picosecond time resolution. We measure the heating and cooling, all within picoseconds, of the electrons and find that, with our sub-picosecond pulses, the highest apparent temperature is reached \SI{0.6}{\ps} before the maximum change in magnitude of the extinction signal.
\end{abstract}

\maketitle

\section{Introduction}

\subsection{Background}

Plasmonic nanoparticles are remarkable in their ability to couple to light whose wavelength is significantly larger than their size. This property is primarily due to their localized surface plasmon resonance---a resonance of collective electron oscillations governed by the spatial confinement of the electron gas \cite{maier2007plasmonics}.


Leaving aside the rich physics created by various nanoparticle geometries, from tunable spectroscopic properties to powerful near-field effects, absorption of light by a plasmonic nanosphere can be understood in the following simplified way:

If the dielectric permittivity of the material is known, the key parameters of scattering and absorption can be calculated from Maxwell's equations (for a sphere, using Mie theory \cite{bohren2008absorption}). For far-field interactions, these are conveniently given in terms of cross sections: the scattering cross section $\sigma_\mathrm{sca}$, the absorption cross section $\sigma_\mathrm{abs}$, and the extinction cross section $\sigma_\mathrm{ext} = \sigma_\mathrm{sca} + \sigma_\mathrm{abs}$. Through $\sigma_\mathrm{abs}$, we have access to the rate at which energy is absorbed by the nanoparticle, and thus to the amount of energy that must (eventually) leave the nanoparticle when it equilibrates with its environment.

The vast majority of the energy absorbed is converted to heat, which the (now hotter) nanoparticle will dissipate \cite{Dubi-Sivan}. A vanishingly small fraction of the energy absorbed goes towards radiative emission channels referred to as photoluminescence; while this emission is measurable and useful---and indeed central to the technique used in this work---for the purposes of the energy balance, it can safely be neglected. 

This simple picture describes low-intensity near-steady-state continuous-wave excitation; however, it does not capture the dynamics of light absorption or the ultrafast heat transfer dynamics which occur in the metal. These come to the fore in particular when a plasmonic nanoparticle is excited with a pulsed laser with a pulse width below about a picosecond.

Ultrafast laser absorption in a metal, be it in a nanoparticle, a film, or bulk metal, is typically modeled using a two-temperature model\cite{anisimov_electron_1974,voisin_ultrafast_2001,hodak_spectroscopy_1998,delFatti_nonequilib_2000}; the electron gas and the metal lattice are treated as two distinct coupled subsystems with independent temperatures. In bulk metal, both temperatures must be treated as temperature fields varying in space; in these cases, heat diffusion can play a critical role in the dynamics \cite{block_tracking_2019,sivan_ultrafast_2019}. In nanoparticles that are sufficiently small compared to the wavelength and focus size of the heating laser, the heating of the particle is uniform \cite{thermo-plasmonics-basics,un_sivan_jap19}. The temperatures of the electrons and of the lattice can then be taken to be constant across the entire volume of the particle, and may be treated as scalars.

In the two-temperature model, under ultrafast laser excitation, we typically ignore the initial heating and thermalization, and describe the dynamics from the point in which the electron system has fully self-thermalized to a Fermi--Dirac distribution. This neglects any heating of the lattice before that point. When the excitation pulse width is of the same order as the thermalization time of the electrons (\SI{\sim0.5}{\ps} in gold \cite{groeneveld_femtosecond_1995, fann_electron_1992,voisin_ultrafast_2001, Aeschliman_e_photoemission_review}), the fact that the electron distribution is instantaneously non-thermal is assumed to barely have an impact on the measurement. Assuming instantaneous thermalization, we model the evolution of the temperatures using the following coupled ordinary differential equations \cite{hodak_spectroscopy_1998,brown_ab_2016,brown_experimental_2017}:
\begin{subequations}\label{eqs:ttm}
\begin{align}
C_e(T_e)\frac{\mathrm d T_e}{\mathrm d t} & = -G (T_e - T_l) \label{eq:ttm_e} \\
C_l \frac{\mathrm d T_l}{\mathrm d t} & = G (T_e - T_l) - Q_\mathrm{bath}(T_l) \label{eq:ttm_l}
\end{align}
\end{subequations}
where $T_e$ and $T_l$ are the temperatures of the electrons and the lattice, $C_e$ and $C_l$ are the corresponding heat capacities, $G$ is an electron--phonon coupling constant, and $Q_\mathrm{bath}$ represents heat transfer from the lattice to the environment. Direct heat transfer from the electrons to the environment is neglected. Note that the electron heat capacity depends on the electron temperature. The electron excitation generally relaxes within a few picoseconds, while the lattice cools down over a longer period of time, depending on the thermal properties of the environment, especially the thermal conductivity and, for small particles in particular, the interfacial (Kapitza) thermal resistance. Under typical conditions, relaxation of the lattice takes place over (at least) a period of hundreds of picoseconds.

Because of the low heat capacity of the electrons (viz. about \SI{2e-2}{\joule\per\cubic\centi\meter\per\kelvin}, \cite{hodak_spectroscopy_1998} or \SI{0.8}{\percent} of that of the gold lattice, at standard temperature and pressure), the instantaneous temperature of the electrons upon pulsed excitation can become very high. It has been estimated by different authors under various conditions to reach into the thousands of kelvin under modest illumination, both indirectly from extinction and/or scattering measurements \cite{hodak_spectroscopy_1998,schoenlein_femtosecond_1987,elsayed-ali_femtosecond_1991,brown_experimental_2017}, and more directly, using anti-Stokes emission.\cite{huang_resonant_2014}

The two-temperature model in eq \oureqref{eqs:ttm}, by allocating the electrons to a Fermi distribution with a certain elevated temperature $T_e$ at time zero, does not take into account how the electrons got there. Considering this is important for two main reasons: Firstly, if the excitation laser pulse width is comparable to the electron--phonon coupling time, then they should be considered together in order to obtain accurate predictions for the temperature dynamics. Secondly, optical excitation of the electron gas does not yield a thermal distribution of electrons, and this must be considered if the duration of the excitation laser pulse approaches the relaxation time of the electron distribution. Indeed, several studies have considered non-thermal electron distributions and how they thermalize.\cite{sun_femtosecond-tunable_1994,groeneveld_femtosecond_1995,voisin_ultrafast_2001,carpene_ultrafast_2006,Italians_hot_es,labouret_nonthermal_2016,saavedra_hot-electron_2016,Aeschliman_e_photoemission_review}

\subsection{Anti-Stokes emission as a measure of temperature\label{sec:antistokes}}

When a nanoparticle is illuminated at a suitable wavelength, it can be detected optically, either through elastic (Rayleigh) scattering, in which the detected photons have the same energy as the illumination, or through inelastic processes, in which the detected photons have either gained energy from or lost energy to the nanoparticle. If the emitted photons have a higher energy (blue-shift), this is known as an anti-Stokes process, if they have a lower energy (red-shift), a Stokes process.

The simplest such inelastic process would be a single scattering event, in which a single incoming photon interacts once with a single particle (such as an electron, a hole, or a phonon) in the material, exchanges some energy and momentum with it, and is reemitted. Such lowest-order processes are collectively known as Raman scattering. A wide variety of higher-order processes may also occur: an interaction may involve two or more photons, two or more electrons, holes or phonons, or some combination. Looking at such processes from a different perspective, the photon or photons may create an excited state, and this excitation may then decay radiatively some time later after some number of interactions with electrons, holes, and phonons.

All this is general enough to be true for any sample which interacts with light, but which types of interactions will dominate varies considerably. The main variable here is the availability of excited states: in insulators or semiconductors with a band gap greater than the photon energy, any excited states are negligible and the inelastic interactions are dominated by the lowest order process, Raman scattering. In molecules and semiconductors which happen to have long-lived excited states resonant with the incoming light, these, instead, dominate the inelastic emission. This is known as fluorescence.

In metal nanoparticles specifically, excited states are available in the conduction band (which we may think of, approximately, as electron-hole pairs, and, collectively, localized surface plasmon polaritons), but they have relatively short lifetimes on  the order of tens of femtoseconds \cite{link_spectral_1999}. This means that their populations remain small at all times and that the excited charge carriers only experience a limited number of interactions with the thermal bath. In other words, the excited states decay well before thermal equilibrium is achieved---if they decay radiatively, this is known as \enquote{hot} luminescence \cite{klein_equivalence_1973}.

In general, hot luminescence in metals includes Raman scattering as a first-order contribution \cite{shen_distinction_1974,cai_anti-stokes_2019,mertens_how_2017} and there is no easy way to separate Raman scattering from higher order contributions. In any case, energy is gained and lost through interactions with the electron and phonon subsystems. 
Thus, anti-Stokes emission, in which the emitted photons have an energy $\hbar (\omega_L + \delta \omega)$, which is larger than the energy $\hbar \omega_L$ of the laser illumination, can only occur when the metal has occupied states with an energy of $\epsilon = \hbar \delta\omega$ from which provide the gained energy. 
Both phonon and electron states may, in principle, contribute, to these processes; this is reflected in the terminology of Raman scattering: \enquote{traditional} Raman scattering exchanges energy with vibrational states, while electronic Raman scattering is universally couched in terms of scattering with quasi-free electrons, which dominate the behavior of metals.

The emission spectrum will depend on the energy distribution of the states which might donate energy, $n(T,\epsilon)$. Broadly speaking, the emission spectra will follow a Boltzmann-type distribution: If electron--photon interactions dominate, which is likely at the very least at large $\delta\omega$ due to the higher temperature achieved by the electrons, it should obey Fermi--Dirac statistics. If, on the other hand, interactions with phonons dominate, it should follow Bose--Einstein statistics. In practice, the differences between the predictions of the three distributions are marginal for the range of $\delta\omega$ which we have access to experimentally (Nota bene, other authors have made the same observation about their measurements. \cite{cai_anti-stokes_2019}). The emission can then be described by an effective  electron temperature which matches up well with \emph{a priori} predictions \cite{cai_anti-stokes_2019,crampton_junction_2018} and, by extrapolating to zero laser heating, to measure the temperature of the environment  \cite{carattino_gold_2018}.

In the following, the term \enquote{photoluminescence} will be used for inelastic emission in a broad sense, without prejudice as to the mechanism which causes it.

\section{Method}
\subsection{Premise}
\label{sec:premise}

We probe the dynamics which occur in response to pulsed irradiation of a single gold nanosphere using time-resolved anti-Stokes spectroscopy. Studying a \emph{single} nanoparticle rather than an ensemble is valuable not just in order to account for heterogeneity of real nanoparticles, but, in any experiment that involves heating, in order to avoid cumulative heating of multiple particles, which can be considerable \cite{baffou2013photoinduced,sivan_thermal_2019}. To achieve ultrafast time resolution, we use a two-color pump-probe technique:

Two $\sim$ \SI{350}{\femto\second} laser pulses are sent to the sample with a particular delay $\tau$ between one and the other. The first, the \enquote{pump} pulse, with a central wavelength of $\lambda = \SI{785}{\nano\meter}$, is far to the red of the localized surface plasmon resonance of the nanosphere, meaning it will not contribute to our photoluminescence measurement in the visible. The second, the \enquote{probe} pulse, at $\lambda = \SI{594}{\nano\meter}$, is near the resonance. We acquire emission spectra in the neighborhood of \SI{594}{\nano\meter} using spectral filters to remove the signal from the lasers and from elastic scattering.

Both colors are absorbed by the nanosphere, though the pump beam is absorbed significantly less efficiently. What's crucial is that since the pump laser is ca.\ \SI{0.5}{\eV} to the red of both our observation range and the plasmon resonance, (i) any possible emission is not enhanced by the plasmon, and (ii) whatever inelastic emission may be detectable without plasmon enhancement is spectrally separated from the probe signal. In other words, the pump is, on its own, effectively invisible in our measurement.

Note that the following is a heuristic description of the physics involved with the goal of a rough understanding of our measurements, and should not be understood as a robust theoretical model.

The anti-Stokes emission due to the probe pulse is non-trivial function $n(t, \epsilon, I_\mathrm{vis}, I_\mathrm{ir}, \dots)$ of the of the total electron distribution $f(t, \epsilon, I_\mathrm{vis}, I_\mathrm{ir}, \dots)$,
which can be expanded in terms of the contributions of the two (scalar) laser intensities as
\begin{align}
n(t, \epsilon, I_\mathrm{pr}, I_\mathrm{pu}, \dots) & \approx 
       n_0(\epsilon)
     + \partial_\mathrm{pr} n(t - \tau, \epsilon) I_\mathrm{pr} \nonumber \\
    & + \partial_\mathrm{pu} n(t, \epsilon) I_\mathrm{pu}  + \cdots 
    \label{eqn:pop_of_time}
%
%
\end{align}
where $\partial_\mathrm{pr} \equiv (\partial/\partial I_\mathrm{pr})_{I_\mathrm{pu}}$ and vice versa,
if we expand it to first order in terms of the contributions of the two (scalar) laser intensities: the first-order contribution $\partial_\mathrm{pu} n I_\mathrm{pu}$ of the pump pulse, arriving at time zero, and the first-order contribution $\partial_\mathrm{pr} n I_\mathrm{pr}$ of the probe pulse, arriving a time $\tau$ later. This approximation is valid for small perturbations, where the higher-order contributions vanish and the effects of the two pulses can thus be described as linear and additive.


Given that the anti-Stokes emission is characteristic of $n(t, \epsilon, \dots)$, we can write, regardless of the underlying mechanism,
\begin{align}
I_\mathrm{AS}(\hbar\omega) = & \int\limits dt I^{t}_\mathrm{pr}(t) f(\hbar\omega) g(\epsilon) n(t, \epsilon, I_\mathrm{pr}, I_\mathrm{pu}, \dots) \nonumber\\
 & + \mathcal{O}(n^2)
 \label{eqn:pl-integral}
\end{align}
where $\epsilon = \hbar\delta\omega =  \hbar\omega - \hbar\omega_\mathrm{pr}$, $I^{t}_\mathrm{pr}(t) \propto I_\mathrm{pr}$ is the time-dependent intensity of the probe laser, $\hbar\omega_\mathrm{pr}$ is its photon energy, $g(\epsilon)$ represents a density of states, and $f(\hbar\omega)$ represents a probability of emission. If we now approximate the time-dependence of $I^{t}_\mathrm{pr}(t)$ as a Dirac $\delta$ function with the peak at $t = \tau$, we get
\begin{equation*}
I_\mathrm{AS}(\hbar\omega) \approx f(\hbar\omega) g(\epsilon) n(\tau, \epsilon, \dots) \times I_\mathrm{pr}
\end{equation*}
which, using the approximation from eq \oureqref{eqn:pop_of_time}, becomes
\begin{align}
I_{\mathrm{AS}}(\hbar\omega) & \approx f(\hbar\omega)g(\epsilon) \times \Big[ n_0(\epsilon) I_\mathrm{pr} \nonumber \\ 
    & + \partial_\mathrm{pr} n (t=0,\epsilon) \times {I_\mathrm{pr}}^2 \nonumber\\
    & + \partial_\mathrm{pu}n (t=\tau,\epsilon) \times I_\mathrm{pr} \times I_\mathrm{pu} \Big] = \nonumber\\
    & = I_\mathrm{AS}^{(0)} + \Delta I_\mathrm{AS}^\mathrm{pr} + \Delta I_\mathrm{AS}^\mathrm{pu}
\label{eqn:aS-taylor}
\end{align}
where $\delta\omega=\hbar\omega-\hbar\omega_\mathrm{pr}$. The first term is a constant background, the second term accounts for the effect of the probe pulse alone (independent of $\tau$) and the third term is responsible for the pump--probe signal due to the combination of both laser pulses. Note that the constant background is small compared to the other terms due to the comparatively low unperturbed electron temperature.

\begin{figure}
\includegraphics{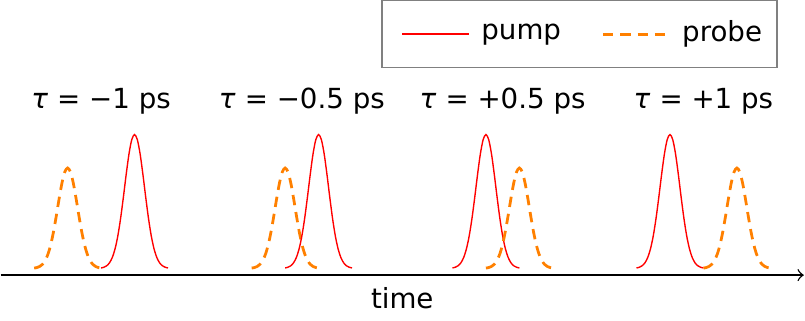}
\caption{
    Visual representation of different pulse delays $\tau$. Note that, typically, $\tau \ge 0$ for the pump pulse to have an effect on the measurement with the probe.
    \label{fig:pulses}
}
\end{figure}

By varying the delay $\tau$ between the pump and probe pulses, as sketched in Fig.~\ref{fig:pulses}, and examining the anti-Stokes spectra due to the probe, we can then study the
the dynamics of $\Delta I_\mathrm{AS}^\mathrm{pu}$, and with it the dynamics of $\partial_\mathrm{pu}n (\tau,\epsilon)$ and of the electron population, with sub-picosecond time resolution.

\subsection{Experimental details}

Two correlated laser pulses are prepared using a titanium-sapphire (Ti:Sapph) laser, which produces a \SI{75.8}{\MHz} train of near-transform-limited pulses with a central wavelength of \SI{785}{\nm} (the \enquote{pump} pulses), and with a frequency-doubled optical parametric oscillator (OPO), which is pumped by the Ti:Sapph laser. The fact that one laser pumps the other means that every pulse from the OPO corresponds to a Ti:Sapph pulse---the two are locked together. The visible output of the OPO is tuned to \SI{594}{\nm} (the \enquote{probe}). Both pulses individually pass suitable dielectric band-pass filters which are well-matched to the notch filters in the detection. Additionally, we use a solid state continuous-wave \SI{532}{\nm} laser to identify particles, for fine adjustments, and for CW photoluminescence spectra.

The pump pulse width is measured to be \SI{350}{\fs} after an acousto-optic modulator in the beam path, which considerably lengthens the pulse. This component is needed only for the acquisition of transient extinction timetraces (as shown, e.g., by \citet{ruijgrok_damping_2012}). Further optical components other than the objective should have little to no further effect on the pulse width; we can estimate the effect of the objective by approximating it as a combination of BK7 glass and fussed silica that adds to a total length of \SI{\sim5}{\cm}, resulting in a final pulse width at the sample of approximately \SI{500}{\fs} (at \SI{785}{\nm}).

The delay $\tau$ between the pulses is adjusted using an optomechanical delay line with a length of up to \SI{1}{\ns} in the path of the near-infrared (pump) beam. All three are then carefully overlapped in space to follow the same optical path and tightly focussed on the sample with an oil-immersion objective (Olympus, $\mathrm{NA} = 1.4$). A second objective on the far side of the sample (Olympus, $\mathrm{NA} = 0.75$) collects the transmitted light, which passes a spectral filter to remove the NIR component before being focussed on a fast photodiode (FEMTO Messtechnik). This arrangement allows measuring the change in extinction of the visible pulse as a function of inter-pulse delay and the acquisition of transient extinction timetraces (see \citet{ruijgrok_damping_2012}), but is further not essential for the work discussed here. The spatial overlap of the pulse trains is optimized using pump-probe extinction contrast (for the two pulses) and photothermal contrast (to overlap the pulses with the CW laser).

\begin{figure}
\includegraphics{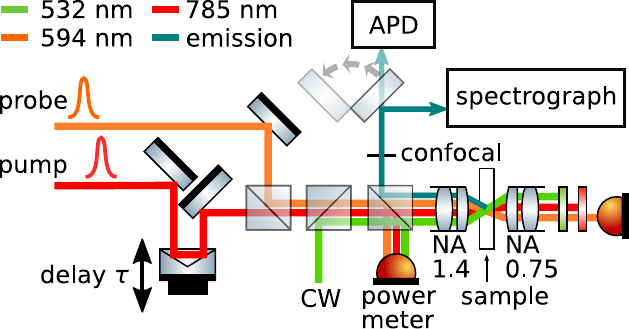}
\caption{Highly simplified sketch of the experiment showing the logical beam paths.
\label{fig:antistokes-setup-sketch}
}
\end{figure}

Meanwhile on the near side of the sample, the light (back-scattering and photoluminescence) collected by the $\mathrm{NA} = 1.4$ objective is split off into the confocal detection path with a 50:50 beam splitter. After passing through a confocal pinhole (\SI{50}{\um}) and a series of notch filters for both colors, the emitted light is sent either to an avalanche photodiode, used for focussing, or to a liquid N$_2$-cooled spectrograph (Acton Research SpectraPro-500i). Fig.\ \ref{fig:antistokes-setup-sketch} shows a rough logical sketch of the beam paths from source to detection, disregarding most optical components.

Photoluminescence spectra are acquired with an integration time of \SI{240}{\second}, with the exception of some spectra recorded at very low or very high intensity. The very long integration time is necessitated by the low intensity of the signal from a single nanoparticle; in other work measuring picosecond-timescale behavior of single nanoparticles, the need for such long integration times is often obviated by the use of lock-in and/or interferometric measurements instead of photoluminescence. All spectra shown here have been normalized by the integration time. Before the acquisition of each spectrum, the particle is brought into focus by maximizing photoluminescence with an automatic focussing routine. This is done to compensate for any slight drift in the system that may occur over the course of a long measurement.

During the acquisition, the inter-pulse delay $\tau$ is kept constant. Note that the spectra are integrated over billions of full cycles; as shown in eq~\oureqref{eqn:pl-integral}, the measured photoluminescence is the integral (for one value of $\tau$) of the instantaneous photoluminescence, which depends on the instantaneous incident intensity and the instantaneous electron distribution. The instantaneous photoluminescence at the peak (in time) of the probe pulse contributes more strongly to the measurement than the tails of the pulse.

Quoted pulse energies are measured as average powers in the back focal plane, before the objective, as indicated in Fig.\ \ref{fig:antistokes-setup-sketch}. They are not normalized by the transmission of the objective or by the absorptivity of the sample.

The sample consists of \SI{100}{\nm} gold nanospheres (Nanopartz Inc.) spin-coated on a glass cover slip (Menzel). Spheres were chosen as other geometries are more liable to reshape under pulsed illumination. The nanoparticles are very dilute; in all cases the nearest neighbor of the nanoparticle being studied was more than \SI{\sim 2}{\um} away. The nanoparticles on the glass are further immersed in a reservoir of ultrapure water.

\section{Results}
\subsection{Dependence of the electron effective temperature on intensity}

\begin{figure*}[t!]
\includegraphics{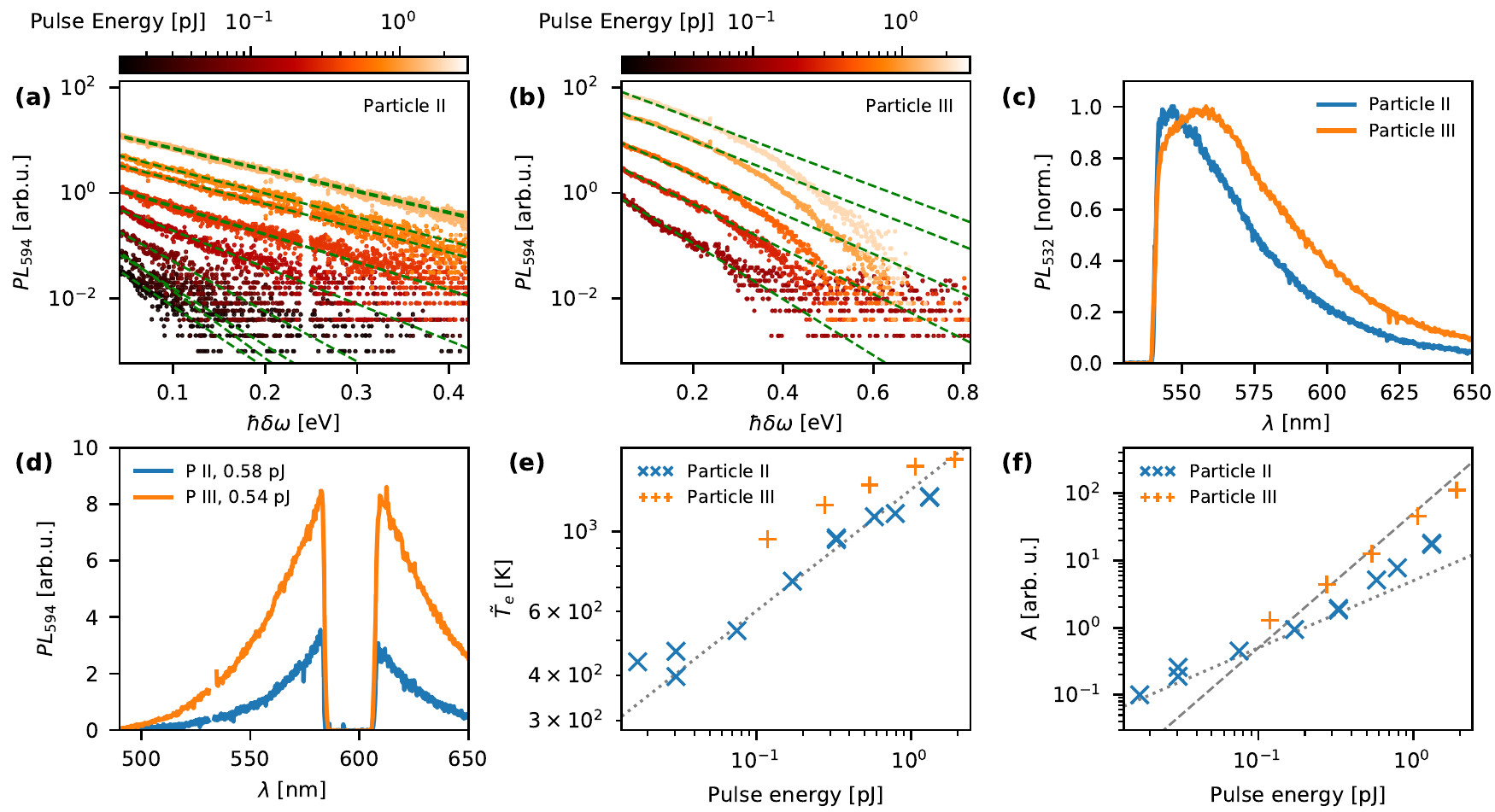}
\caption{
    Anti-Stokes spectra (probe only) of particles \enquote{II} \subfigid{(a)} and \enquote{III} \subfigid{(b)} excited by \SI{594}{\nm} pulses; different colors represent different pulse energies. Dashed lines are fits to the Boltzmann distribution, eq~\oureqref{eqn:spectra_fit}. \subfigid{(c)} Photoluminescence spectra of the same particles excited at \SI{532}{\nm} (CW). \subfigid{(d)} Examples of spectra including the Stokes and anti-Stokes components, on a linear scale. \subfigid{(e)} Temperatures derived from the fits. \subfigid{(f)} Scaling factors $A$ derived from the fits. \subfigid{(e)} and \subfigid{(f)} are plotted on a log--log scale; the dotted grey lines have a slope of 1, the dashed grey line in (f) has a slope of 2. The grey lines are not fits to the data. The pulse energies quoted are measured in the back focal plane; note that the particles absorb only a fraction of the available energy.
\label{fig:powerdep_spectra}}
\end{figure*}

In order to establish clearly which parts of the resulting anti-Stokes spectra are due to heating by the probe pulse---in the language of eqs \oureqref{eqn:pop_of_time} and \oureqref{eqn:aS-taylor}, in order to get an idea of $n_0(\epsilon) + \partial_\mathrm{pr} n(\epsilon)$---we first measure Stokes and anti-Stokes spectra of the nanoparticle using the probe alone, at different excitation powers. The spectra are shown in Fig.\ \ref{fig:powerdep_spectra}.

In general terms, if the distribution $n(\epsilon)$ can be approximated with a Boltzmann-form distribution, we can say that the intensity for the probe-only experiment should follow
\begin{equation*}
I_\mathrm{AS}(\hbar\omega) \propto f(\hbar\omega)g(\hbar\delta\omega)\exp\left( \frac{-\hbar\delta\omega}{k_B \tilde{T_\mathrm{e}}} \right)
\end{equation*}
where we identify the effective electron temperature $\tilde{T_\mathrm{e}}$ with the effective electron temperature of the two-temperature model (eq~\oureqref{eq:ttm_e}). The use of a Boltzmann factor here disguises the distinction between the electron and phonon thermal baths, and is only justifiable if $\hbar\delta\omega$ is sufficiently large, i.e. far from the laser. In our measurements, the spectral filter blocking out the laser largely obscures the energy range in which this approximation may break down.


\citet{carattino_gold_2018} could make two simplifying assumptions: firstly, since they were using nanorods rather than nanospheres, they could identify $f(\hbar\omega)$ with a sharp plasmon resonance; this approximation is not valid for spheres as their resonance is much broader. Further, they were operating far enough from the main interband transitions of gold to assume $g(\hbar\delta\omega)$ is constant. As the resonance of gold nanospheres is much closer to the interband transitions, this approximation is not valid, either \cite{yorulmaz_luminescence_2012}. Their use of the surface plasmon resonance as a normalizing factor can therefore not be replicated here; this is consistent with the lack of apparent signature of the plasmon in the spectra in Fig.\ \ref{fig:powerdep_spectra}.

We resort instead to a rather simpler approximation: we assume that the function \(f(\hbar\omega)g(\hbar\delta\omega)\) varies much more slowly than the exponential Boltzmann factor. This leaves us with
\begin{equation}
I_\mathrm{AS}(\hbar\omega) = A\,\exp\left( \frac{-\hbar\delta\omega}{k_B \tilde{T_\mathrm{e}}} \right)
\label{eqn:spectra_fit}
\end{equation}
where $A$ is a proportionality coefficient and $\tilde{T_\mathrm{e}}$ is the effective extracted temperature of the electrons. 
Note that for relating this effective temperature to the real electron temperature a more detailed 
understanding of the mechanism for anti-Stokes emission is needed.

Due to the lack of a \emph{sharp} resonance in a nanosphere, this approximation is reasonable sufficiently far from the interband transitions. The nearby interband transition at \SI{470}{\nm} \cite{etchegoin_analytic_2006} corresponds to $\hbar\delta\omega \approx \SI{0.5}{\eV}$ (i.e. $\SI{6e3}{\kelvin}\times k_B$). Bearing in mind the broadness of the interband transition for a small particle at high (i.e., room) temperature, this corresponds to the region in Fig.\ \ref{fig:powerdep_spectra}b where we clearly see the fits to eq~\oureqref{eqn:spectra_fit} break down.

The same approximation has been used successfully by other authors in the past, such as \citet{he_surface_2014} and \citet{cai_anti-stokes_2019}. Authors inclined to interpret similar measurements as electronic Raman scattering, such as \citet{crampton_junction_2018}, instead assume that $f(\hbar\omega)g(\hbar\delta\omega) \propto \omega^3$, which, in practice, is equivalent to our approximation since $\omega^3$ varies much more slowly than the Boltzmann factor.

This simplistic Boltzmann description fits the data well up to $\hbar\delta\omega \approx \SI{0.4}{\eV}$, at which point the approximations start to break down as expected. Note, however, that the extracted temperatures reflect the slope (of the logarithm) in the region where the fit is good ($\delta\omega \lesssim \SI{0.4}{\eV}$), and are thus not particularly affected by the interband transitions.

The effective temperatures $\tilde{T_\mathrm{e}}$ and amplitudes $A$, shown in Fig.\ \ref{fig:powerdep_spectra}e and \ref{fig:powerdep_spectra}f respectively, broadly show the expected features:

(i) The effective temperatures reach \SI{>1000}{\kelvin}, and increase monotonically with increased incident power. The high temperatures are consistent with the prediction of electron temperatures that are much higher than the temperature of the gold lattice, and confirm that our measurement is more indicative of the temperature of the electrons than of the lattice.

(ii) The intensity of the anti-Stokes emission also increases with incident power, both due to the increased number of photons at higher powers, and due to the larger temperature achieved by the system. The combination of the two leads to a faster-than-linear increase in $A$ (Fig.\ \ref{fig:powerdep_spectra}f), as expected from eq~\oureqref{eqn:aS-taylor}.

(iii) The rate of temperature change with heating power $\partial T / \partial P$ slows with increased power. Multiple factors contribute to this remarkable effect: On the one hand, the heat capacity of the electrons $C_e(T_e)$ increases as the temperature goes up, meaning the same amount of heat corresponds to a smaller temperature increase \cite{hodak_spectroscopy_1998}. On the other hand, the permittivity, and thus the amount of absorbed heat, also depends on the temperature\cite{Shalaev_ellipsometry_gold,PT_Shen_ellipsometry_gold}, which has been shown theoretically to lead to a similar slowing of the change in temperature \cite{sivan_nonlinear_2017,gurwich_metal_2017}.

(iv) The effective temperature roughly approaches room temperature as $P \rightarrow 0$. This is, however, far from exact for two main reasons. Firstly, although the dependence should be linear at low power, how low the cut-off power for linearization should be is difficult to say without greater knowledge of the nonlinear function $T(P)$, and secondly, the margins of error are substantial: on the one hand, the rather low signal-to-noise ratio at the lowest powers increases the uncertainty as the temperature decreases. 

Evidently, a more detailed model of the emission mechanism as well as of the thermodynamic and photothermal effects in this system is needed to accurately calculate and understand the true electron temperature, and its behavior as a function of power.

Fig.\ \ref{fig:powerdep_spectra} shows a remarkable heterogeneity between the two particles: particle III appears to be heated significantly more efficiently than is particle II, even though the particles have nearly the same size; this implicates other factors that may influence the absorption at \SI{594}{\nm} and elsewhere, such as the precise shape of the particles (including surface facets), interactions with the substrate, or crystal defects \cite{setoura2013observation}. This heterogeneity demonstrates the importance of studying single gold nanoparticles rather than relying on ensemble measurements.


\subsection{Emission dynamics}
\label{sec:temp-dyn}

\begin{figure*}
\includegraphics{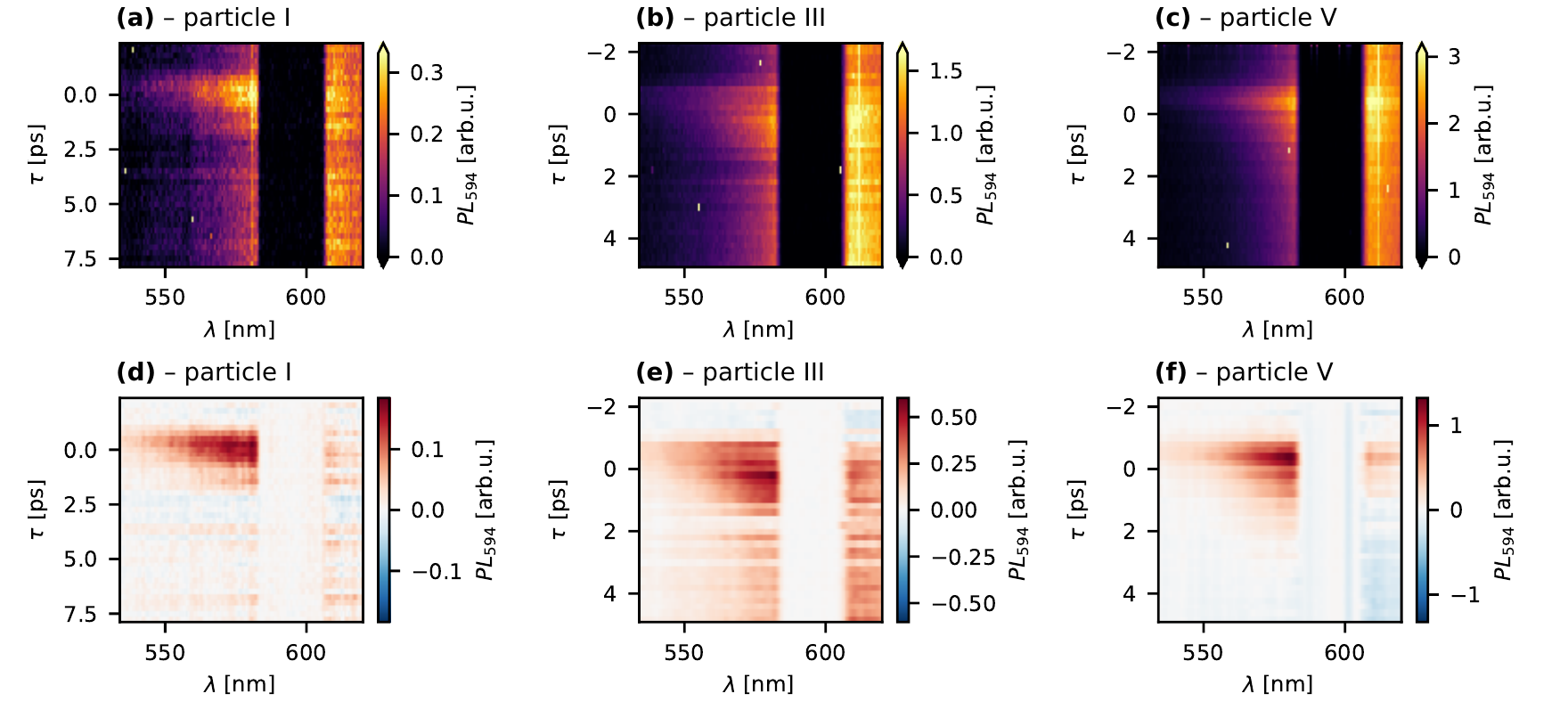}
\caption{Time-resolved spectra of three different \SI{100}{\nm} diameter (nominal) gold nanospheres, as a function of the wavelength $\lambda$ and pump-probe inter-pulse delay $\tau$. \subfigid{(a--c)} The raw spectra on the top, \subfigid{(d--f)}, spectra with the $\tau$-independent contribution of the probe pulse subtracted on the bottom, showing only the contribution of the pump pulse $\Delta I_\mathrm{A}^\mathrm{pu}$.
Probe pulse energies were
\subfigid{(a,d)} \SI{85(3)}{\fJ},
\subfigid{(b,e)} \SI{119(11)}{\fJ} and
\subfigid{(c,f)} \SI{190(11)}{\fJ}. The units on the color scales are arbitrary but all equivalent to each other. $\tau = 0$ is defined as the peak of the pump-probe extinction spectra as shown in Fig.\ \ref{fig:twocolour_pars}.
\label{fig:twocolour_spectra}}
\end{figure*}

In the two-color experiment, spectra are recorded while focusing two pulse trains on the particles, a probe pulse which causes the measured emission, and a pump pulse, whose direct contribution to the measured anti-Stokes spectra is negligible. In the language of equation \ref{eqn:aS-taylor} we now measure the three terms together. The delay between the pulses $\tau$ is varied from one spectrum acquisition to the next. $\tau$ is calibrated such that at positive $\tau$ the probe pulse arrives after the pump pulse, and such that $\tau = 0$ corresponds to the initial maximum of a \enquote{traditional} pump-probe extinction timetrace.

Fig.\ \ref{fig:twocolour_spectra} shows such time-resolved spectra for three different \SI{100}{\nm} (nominal) gold nanospheres. The effect of the presence of the pump pulse on the spectra is limited to a period of \SI{\sim 1}{\ps}; at longer delays the emission returns to its initial state.

The change on the anti-Stokes side of the spectrum is significantly greater than the change on the Stokes side. This aligns with our expectations, as the anti-Stokes emission has a stronger dependence on the temperature than the Stokes emission; changes in Stokes emission due to elevated electron temperatures are a higher-order effect which we neglect in this work. In future work, it may be desirable to account for such higher-order terms and for secondary effects, such as the change in the dielectric constant.

As stated in Sec. \ref{sec:premise}, the measured anti-Stokes spectra are characteristic of the total electron distribution when the probe pulse arrives, as given in eq~\oureqref{eqn:pop_of_time}. Here, two approaches for analysis present themselves:

For a straightforward parametrization, we can simply fit eq~\oureqref{eqn:spectra_fit} to the spectra for each time $\tau$. The equation, now
\begin{equation}
I_\mathrm{AS}(\tau, \hbar \delta\omega) = A(\tau)\,\exp\left(\frac{-\hbar\delta\omega}{k_B\,\tilde{T_\mathrm{e}} (\tau)}\right)
\label{eqn:tdep_spectra_fit}
\end{equation}
immediately gives us two time-dependent parameters that characterize the resulting emission quite well: an effective temperature $\tilde{T_\mathrm{e}}(\tau)$ and a quasi-amplitude $A(\tau)$.

Alternatively, we can make use of the fact that the effect of the pulses is additive under the approximation of eq~\oureqref{eqn:aS-taylor} and subtract the $\tau$-independent component, which is assumed to be equal to the spectrum at $\tau = \SI{-2.2}{\ps}$: the $\tau$-dependent term must be zero when the pump pulse arrives after the probe pulse for reasons of causality.

Figs.\ \ref{fig:twocolour_spectra}d--\ref{fig:twocolour_spectra}f show these difference spectra, i.e. the term $\Delta I_\mathrm{AS}^\mathrm{pu}$ alone. For the purpose of parametrizing the entirety of the data, this latter approach is less practical since $\Delta I_\mathrm{AS}^\mathrm{pu}(\tau, \hbar\delta\omega)$ is overwhelmed by noise after a picosecond or two. For this reason, we will follow both approaches: the first (eq~\oureqref{eqn:tdep_spectra_fit}) to obtain a full parametrization, and the second ($\Delta I_\mathrm{AS}^\mathrm{pu}$ from eq \oureqref{eqn:aS-taylor} with eq \oureqref{eqn:tdep_delta_spectra_fit} below) for selected spectra only.

\begin{figure*}
\includegraphics{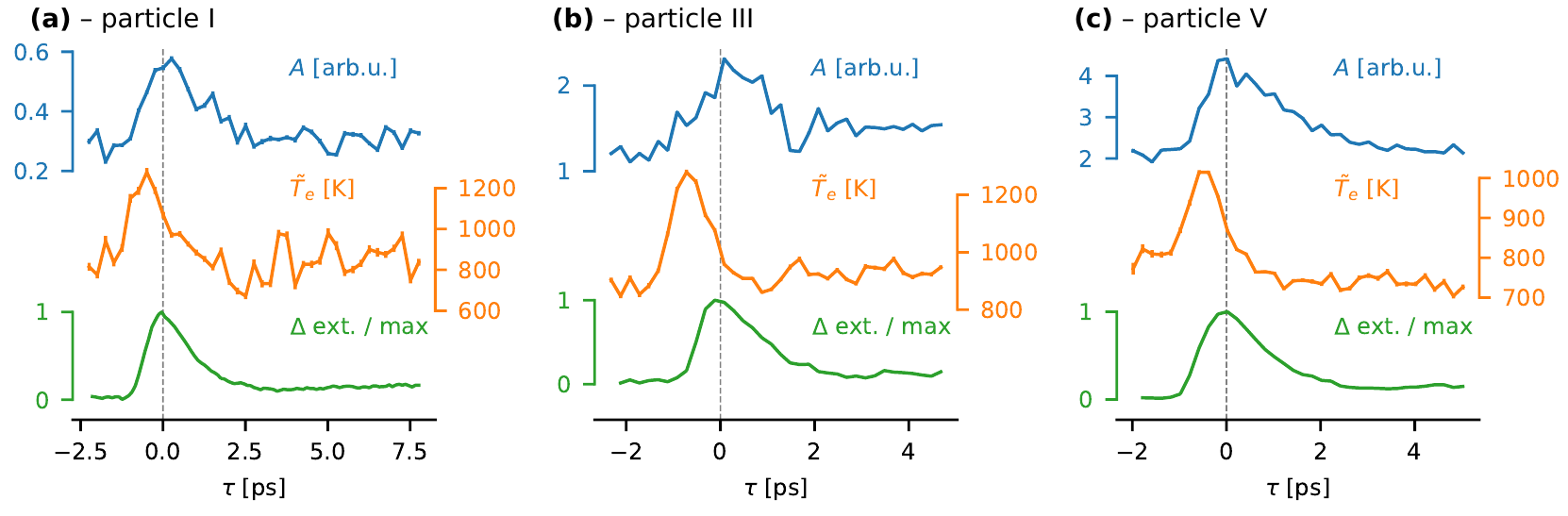}

\caption{Parametrization of the spectra in Fig.\ \ref{fig:twocolour_spectra} in terms of $A(\tau)$ [blue] and $\tilde T_e(\tau)$ [orange] according to eq~\oureqref{eqn:tdep_spectra_fit}. The units for $A(\tau)$ are arbitrary but correspond to the units used in Fig.\ \ref{fig:twocolour_spectra}. Below: pump-probe extinction spectra [green] of the particles for comparison.
\label{fig:twocolour_pars}}
\end{figure*}

The parameters $A(\tau)$ and $\tilde{T_e(\tau)}$ resulting from a fit to eq~\oureqref{eqn:tdep_spectra_fit} are shown in Fig.\ \ref{fig:twocolour_pars}, along with the corresponding pump-probe extinction spectra from which $\tau = 0$ is calibrated. The magnitude of the change in the anti-Stokes spectra, represented by $A(\tau)$ follows the behavior of the change in extinction well: Both have their maxima at the same inter-pulse delay, and both show the same asymmetric behavior as a function of $\tau$, with a steep rising edge as the pulse is absorbed, and a slower \si{\sim\ps} decay as the absorbed energy is released into the metal lattice.

The highest apparent temperature state is, however, reached earlier. Across our measurements, temperature peaked around \SI{0.66 \pm 0.1}{\ps} before the amplitude, which is longer than our pulse width. The high-temperature state then rapidly decays to its initial value as the anti-Stokes intensity increases.

\begin{figure}
\centering
\includegraphics{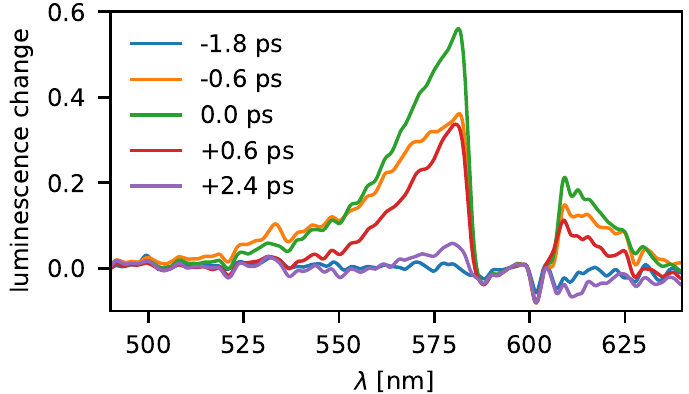}
\caption{Difference spectra from Fig.\ \ref{fig:twocolour_spectra}f: In order of increasing $\tau$, the state before the pump pulse, the highest $\tilde T_e$ state ($\tau = \SI{-0.6}{\ps}$), the highest intensity state ($\tau = 0$), a slightly later state ($\tau = \SI{+0.6}{\ps}$) and a late-$\tau$ state.
\label{fig:individual_as_spectra}}
\end{figure}

This early high-$\tilde T_e$, low-$A$ state is easy to see in the spectra directly now that we know what we're looking for: Fig.\ \ref{fig:individual_as_spectra} shows difference spectra from Fig.\ \ref{fig:twocolour_spectra}f for selected delays $\tau$. The highest-$\tilde T_e$ spectrum, $\tau = \SI{-0.6}{\ps}$, clearly decays much more slowly with decreasing $\lambda$ than does either the tallest spectrum at $\tau = 0$ or indeed any later spectrum, such as the example from $\tau = \SI[retain-explicit-plus]{+0.6}{\ps}$.

Moreover, the spectrum for $\tau = \SI{-0.6}{\ps}$ appears to deviate more strongly from the thermal (exponential) shape than do the others, which may be an indication of a non-thermal \enquote{hot} electron distribution at this early stage. However, the deviation is not clear enough to permit such a conclusion with any degree of certainty.

The fact that the state of maximum temperature occurs so much earlier than the state of peak response in terms of both extinction and luminescence calls into question, at the very least, the implicit assumption that the peak of the pump-probe extinction spectrum, which we refer to as $\tau = 0$, corresponds to the pump and probe pulses arriving at the same moment.

The effective temperature changes due to the pump pulse as reported in Fig.\ \ref{fig:twocolour_pars}, some \SI{300}{\K} or so, appear quite low. This is an artefact caused by the underlying assumption of eq~\oureqref{eqn:tdep_spectra_fit}: for the fit to give a good understanding of the temperatures, the entire electron gas would have to have a well-defined temperature, i.e., would have to be fully thermalized. In terms of eq.~\oureqref{eqn:pop_of_time}, it would require $\partial_\mathrm{pu}n I_\mathrm{pu}$ to have equilibrated with $\partial_\mathrm{pr} n I_\mathrm{pr}$ and $n_0$ before the emission of any photons.

As suggested above, we can get closer to the picture of the electron distribution put forward in eq.~\oureqref{eqn:pop_of_time} by subtracting a baseline spectrum ${I}_\mathrm{pr}(\delta\omega)$ from the full spectrum $I_\mathrm{AS}(\tau, \delta\omega)$ to arrive at the effect of the pump pulse alone, without the effect of the probe pulse. Just as with the full spectrum, we may then imagine it to be caused by a thermal distribution of electrons and fit the spectra to a Boltzmann distribution:
\begin{equation}
\Delta I_\mathrm{AS}^\mathrm{pu}(\tau, \hbar\delta\omega) \overset{\mathrm{fit}}{=} A_\Delta(\tau)\,\exp\left(\frac{-\hbar\delta\omega}{k_B\,T_\Delta(\tau)}\right)
\label{eqn:tdep_delta_spectra_fit}
\end{equation}
where $T_\Delta$ is the effective temperature of the partial electron distribution excited by the pump pulse.

While we expect the initial distribution of electrons to be non-thermal \cite{sun_femtosecond-tunable_1994,voisin_ultrafast_2001,carpene_ultrafast_2006,labouret_nonthermal_2016}, and there is no \emph{a priori} reason why the emission would be thermal, our data does not clearly show it to be otherwise, with the possible exception (as noted above) of the spectrum around $\tau = \SI{-0.6}{\ps}$ in Fig.\ \ref{fig:individual_as_spectra}.

\begin{table}
\centering
\begin{tabular}{c|r|r|r|r}
~ & \multicolumn{2}{c|}{$\tilde{T_e}$ from eq~\oureqref{eqn:tdep_spectra_fit}} %
  & \multicolumn{2}{c}{$T_\Delta$ from eq~\oureqref{eqn:tdep_delta_spectra_fit}} \\
\cline{2-5}
Particle & \multicolumn{1}{c|}{max} & \multicolumn{1}{c|}{$\tau = 0$} %
         & \multicolumn{1}{c|}{max} & \multicolumn{1}{c}{$\tau = 0$} \\
\hline
I & \SI{1278}{\K} & \SI{1067}{\K} & \SI{1661}{\K} & \SI{1206}{\K} \\
III & \SI{1279}{\K} & \SI{958}{\K} & \SI{1756}{\K} & \SI{1004}{\K} \\
V & \SI{1015}{\K} & \SI{868}{\K} & \SI{1306}{\K} & \SI{936}{\K} \\
\end{tabular}

\caption{Apparent temperatures of the total spectra (Figs.\ \ref{fig:twocolour_spectra}a--\ref{fig:twocolour_spectra}c and eq~\oureqref{eqn:tdep_spectra_fit}) and difference spectra (Figs.\ \ref{fig:twocolour_spectra}d--\ref{fig:twocolour_spectra}f and eq~\oureqref{eqn:tdep_delta_spectra_fit}) at maximum temperature and at $\tau = 0$. The values for the former correspond to those in Fig.\ \ref{fig:twocolour_pars}. Note the spectra for particles III and V were measured on the same day and with the same pump pulse energy, viz. ca. \SI{2.2}{\pJ} as measured in the back focal plane.
\label{tab:temps}
}
\end{table}

The results of the fit to eq~\oureqref{eqn:tdep_delta_spectra_fit} are listed in table \ref{tab:temps} alongside the corresponding values (previously shown in Fig.\ \ref{fig:twocolour_pars}) for the fit of the full spectra using eq~\oureqref{eqn:tdep_spectra_fit}. The table shows both the highest temperatures calculated in each case next to the temperature calculated for $\tau = 0$. It shows that the distributions created by the pump pulse initially have significantly higher characteristic temperatures than those due to the probe pulse, which decays rapidly, as seen in the previous figures. We systematically obtained higher apparent temperature values for the subtracted spectra, which may indicate that the pump-excited contribution to the electron distribution is indeed (as expected) not fully thermalized with the distribution as a whole.

\section{Discussion and conclusion}

We have measured anti-Stokes photoluminescence of single gold nanoparticles due to pulsed illumination with characteristic temperatures of order \SI{e3}{\kelvin}. The temperatures are extracted using a simple Boltzmann approximation (eq~\oureqref{eqn:spectra_fit}) that fits well for intermediate anti-Stokes shifts, i.e., anti-Stokes shifts that are not so small that the Boltzmann distribution would cease to apply, and that are not so large that the interband transitions of gold become significant.

The limit of small $\delta\omega$ does not contribute to the measurement as the corresponding light is close enough to the laser to be rejected by our spectral filters. The limit of large $\delta\omega$ barely contributes to the fit as the signal in that region is very weak in the first place. The extracted temperatures reflect the spectral region which the simple approximations are best suited to.

For a more exact extraction of the temperature from the photoluminescence spectra a more thorough model of the origin of gold nanoparticle photoluminescence is needed. Such a model may also provide more insight into the nature of the apparent temperature. 

In eq~\oureqref{eqn:pop_of_time} we assumed that the excited electron populations created by the two laser pulses are independent of one another. However, in Fig. \ref{fig:powerdep_spectra}e, we see that dependence of the anti-Stokes spectra on pulse energy deviates significantly from linearity. This is not, in and of itself, terribly surprising: we know that the heat capacity of the electron gas and the dielectric permittivity both depend on temperature. The probe pulse energies used in the two-color measurements are on the low side (\SI{\sim0.1}{\pJ}, see Fig.\ \ref{fig:twocolour_spectra} caption), so the linear approximation inherent in eq~\oureqref{eqn:pop_of_time} may still be reasonable, but we cannot exclude higher-order interactions between the pulses that would affect the interpretation of these measurements.

We have shown in Sec. \ref{sec:temp-dyn} that the apparent temperature increase due to pulsed excitation decays on a $\sim$\si{\ps} timescale, which agrees with previous measurements of the electron-phonon coupling time. Surprisingly, the peak apparent temperature is reached early in the process (Fig.\ \ref{fig:individual_as_spectra}), \SI{0.6}{\ps} before the peak amplitude. This time is consistent with the thermalization times previously measured in bulk gold using time-resolved photoemission spectroscopy \cite{fann_electron_1992}. We note that what appears to be an early high-temperature state may in fact be the high-energy tail of a non-thermal state: a lower-temperature Fermi-Dirac--like contribution may be obscured by the spectral filter we use when accompanied by a transient non-thermal \enquote{hot electron} contribution (as measured, e.g., by \citet{fann_electron_1992}). Such a \enquote{hot electron} contribution may appear similar to a thermal distribution in our measurement, but ascribing a temperature to it would not be meaningful.

This observation may allow some deeper insight into the thermalization dynamics of the electron gas in a gold nanoparticle during the arrival of short laser pulses, if it were examined using a detailed model of the electron distribution and its evolution as a function of time.

Our results largely appear, however, to be consistent with a thermal distribution of electrons. We expect this is due to two main factors: Firstly, the probe pulse width is of the same order of magnitude as the thermalization time. In short, there is plenty of time during the probe pulse for the electron distribution to approach a thermal one. Secondly, the spectral filter removing the laser obscures small energy shifts, which means we only measure a relatively small part of the energy distribution, and cannot compare the distributions at low and at high energies as extensively as we might like.

Either one of these shortcomings could be improved somewhat, but there is a trade-off between the two: If one wanted to work with a shorter pulse, this pulse would inevitably have a larger bandwidth (at least \SI{0.2}{\eV} for \SI{\sim10}{\fs}, for instance), which would obscure more of the anti-Stokes spectrum. Vice versa, a much narrower filter would soon require the use of longer laser pulses.
Rather, to measure electron temperature dynamics using anti-Stokes emission, our measurements have to be integrated over a longer period of time (as they are here), and we must rely on theory to clear up the details.

We hope future work will explore the effect of the pulse width on the measurement technique we have introduced, and give the results a more thorough theoretical grounding.

It may also be interesting to extend Fig. \ref{fig:powerdep_spectra}e to higher pulse energies to establish whether the temperature saturates under strong illumination before melting of the particle becomes an issue.

\begin{acknowledgments}
This work was supported by the Netherlands Organisation for Scientific Research (NWO/OCW), 
as part of the Frontiers of Nanoscience (NanoFront) program. MC acknowledges the financial support of the Kavli Institute of Nanoscience  Delft through the KIND fellowships program.
\end{acknowledgments}


%

\end{document}